\documentstyle[12pt,cite,doublespace,epsfig]{article}

\newcommand{\nc}{\newcommand}
\nc{\be}{\begin{equation}}
\nc{\ee}{\end{equation}}
\nc{\bea}{\begin{eqnarray}}
\nc{\eea}{\end{eqnarray}}
\nc{\beas}{\begin{eqnarray*}}
\nc{\eeas}{\end{eqnarray*}}
\nc{\noi}{\noindent}
\nc{\non}{\nonumber}
\nc{\s}[1]{\not \! #1}
\nc{\bb}{\bibitem}
\nc{\lf}{\left}
\nc{\r}{\right}
\nc{\mb}[1]{\makebox[#1]{}}
\nc{\pa}{\partial}
\nc{\sA}{\not \! \! A}
\nc{\newsec}[1]{\section{#1}\mb{0.5cm}}
\nc{\h}{\frac{1}{2}}
\nc{\ra}{\rightarrow}
\nc{\la}{\leftarrow}
\def\mathunderaccent#1{\let\theaccent#1\mathpalette\putaccentunder}
\def\putaccentunder#1#2{\oalign{$#1#2$\crcr\hidewidth
\vbox to.2ex{\hbox{$#1\theaccent{}$}\vss}\hidewidth}}

\nc{\ti}{\mathunderaccent\tilde}
\nc{\sq}{\sqrt{q^2}}

\nc{\bfig}{\begin{figure}}
\nc{\efig}{\end{figure}}
\nc{\half}{\frac{1}{2}}
\nc{\qtr}{\frac{1}{4}}
\nc{\halfP}{\frac{P}{2}}
\nc{\qtrP}{\frac{P^2}{4}}
\nc{\metric}[2]{\eta_{#1#2}}
\nc{\dpsq}[2]{\partial_\mu #1 \partial^\mu #2}
\nc{\ul}{\underline}
\nc{\intall}{\int_{-\infty}^{\infty}}
\nc{\inthalf}{\int_{0}^{\infty}}
\nc{\fullK}{\sum_{ch}\int_{\Omega}d\vec{\xi}\inthalf d\gamma
   \frac{\rho_{ch}(\gamma,\vec{\xi})}{\gamma-(a_{ch}q^2+b_{ch}p\cdot q + c_{ch}
   p^2+d_{ch}P^2 +e_{ch}q\cdot P + f_{ch}p\cdot P) - i\epsilon}}
\nc{\Kintch}{\frac{\rho_{ch}(\gamma,\vec{\xi})}{\gamma-(a_{ch}q^2+
   b_{ch}p\cdot q + c_{ch}p^2+d_{ch}P^2 +e_{ch}q\cdot P + f_{ch}p\cdot P) 
   - i\epsilon}}
\nc{\Kint}{\frac{\rho(\gamma,\vec{\xi})}{\gamma-(aq^2+ bp\cdot q + cp^2
   +dP^2 +eq\cdot P + fp\cdot P) - i\epsilon}}
\nc{\solid}[1]{{\cal Y}_{l}^{m}(#1)}
\nc{\solidq}{{\cal Y}_{l}^{m}(\Lambda^{-1}(P)q)}
\nc{\solidp}{{\cal Y}_{l}^{m}(\Lambda^{-1}(P)p)}
\nc{\loopint}{\int\frac{d^4q}{(2\pi)^4 i}}
\nc{\loopintprime}{\int\frac{d^4q'}{(2\pi)^4 i}}
\nc{\propone}[1]{\frac{1}{#1^2-m^2+i\epsilon}}
\nc{\proptwo}[1]{\frac{1}{(#1)^2-m^2+i\epsilon}}
\nc{\propsubone}[2]{\frac{1}{#2^2-m_{#1}^2+i\epsilon}}
\nc{\propsubtwo}[2]{\frac{1}{(#2)^2-m_{#1}^2+i\epsilon}}
\nc{\JH}{\frac{J}{H}}
\nc{\FH}{\frac{F}{H}}
\nc{\zb}{\bar{z}}
\nc{\ab}{\bar{\alpha}}
\nc{\gb}{\bar{\gamma}}
\nc{\at}{\tilde{\alpha}}
\nc{\zt}{\tilde{z}}
\nc{\gt}{\tilde{g}}
\nc{\vertex}[1]{\Gamma^{[l,m]}(#1,P)}
\nc{\ie}{i\epsilon}

\begin{document}
\thispagestyle{empty}
\begin{flushright}
ADP-96-1/T206,
hep-ph/9601364 \\
\vspace{.5cm}
{\em Talk given at Joint Japan-Australia Workshop on 
Quarks, Hadrons \\and Nuclei, Adelaide, South Australia, November 15-24, 1995\\
(To appear in the conference proceedings)}
\end{flushright}

\begin{center}
{\large{\bf Solving the Bethe-Salpeter Equation in Minkowski Space:
Scalar Theories and Beyond}} \\
\vspace{.5cm}
K.\ Kusaka, K.M.\ Simpson and A.G.\ Williams\\
\vspace{.25cm}
{\it
Department of Physics and Mathematical Physics, University of Adelaide,\\
S.Aust 5005, Australia} \\

\begin{abstract}

The Bethe-Salpeter equation (BSE) for bound states in scalar theories
is reformulated and solved in terms of a generalized spectral representation
directly in Minkowski space.
This differs from the conventional approach, where the BSE is solved
in Euclidean space after a Wick rotation.  For all but the lowest-order
(i.e., ladder) approximation to the scattering kernel, the {\it naive}
Wick rotation is invalid.  Our approach generates the vertex
function and Bethe-Salpeter amplitude for the entire allowed range
of momenta, whereas these cannot be directly obtained from the Euclidean
space solution.  Our method is quite
general and can be applied even in cases where the Wick rotation is not
possible.

\end{abstract}

\end{center}
\vfill
\begin{flushleft}
E-mail: {\it 
kkusaka, ksimpson, awilliam@physics.adelaide.edu.au}
\end{flushleft}
\newpage

\newsec{Introduction}
The Bethe-Salpeter Equation (BSE)~\cite{BS} describes the 2-body component of 
bound-state structure relativistically and in the language of Quantum
Field Theory (for an extensive review, see Ref.~\cite{bsrev}; Ref.~\cite
{bib} is an exhaustive list of BSE literature prior to 1988). 
It has applications in, for example, calculation of electromagnetic
form factors of 2-body bound states and relativistic 2-body bound
state spectra and wavefunctions.

BSEs have been solved analytically for separable kernels and for scattering
kernels in the ladder approximation.
Solutions for BSEs have also been obtained for QCD-based models of meson
structure in Euclidean space; 
these solutions must be 
analytically continued back to Minkowski space. 
It is important to note that analytical continuation back to Minkowski space 
from the Euclidean space solution is quite difficult even for the
simple case of constituents interacting via simple particle exchange 
in the ladder approximation to the scattering kernel. 
In particular, BS amplitudes with timelike momenta cannot be 
unambiguously obtained from the Euclidean space solution without solving 
further singular integral equations.
Furthermore, for any BSE with a 
non-ladder scattering kernel and/or with dressed propagators for the
constituent particles, the proper implementation of this procedure (known
as the Wick rotation~\cite {Wick}) itself is highly non-trivial.  For these
two reasons the direct solution of the Minkowski space BSE
is preferable. Here we outline such a 
method for scalar theories, based on
the Perturbation Theoretic Integral Representation (PTIR) of
Nakanishi~\cite{PTIR}.

The PTIR is a generalisation of the spectral 
representation for 2-point Green's functions to $n$-point functions. 
A general $n$-point function may be written as an integral over
a weight distribution, which contains contributions from graphs
at all orders in perturbation theory. As any graph with $n$ fixed external
legs can be written in PTIR form, this must also be true of any sum of
such graphs.
Hence the PTIR for a particular
renormalised $n$-point function is an integral representation of
the corresponding infinite sum of Feynman graphs with $n$ fixed
external legs.

The scalar-scalar BSE has been solved numerically in the ladder approximation
after the Wick rotation by Linden and Mitter~\cite{lm}. Here we present 
Minkowski space solutions to the ladder BSE, which will act as a check of our
implementation of the approach to be used here. Our numerical
solutions are obtained by using the
PTIR to transform the equation for the proper bound-state vertex,
which is an integral equation involving complex distributions, into a
real integral equation. 
This equation may then be solved numerically
for an arbitrary scattering kernel~\cite{kw}. We will restrict 
our consideration of explicit numerical solutions
to the ladder approximation, although the approach is a 
completely general one. 
Calculations for non-ladder kernels are under way and these results will
be presented elsewhere~\cite{kkt}.
As a specific example of a scalar theory to
which our formalism may be applied, consider the $\phi^2\sigma$ model,
which has a Lagrangian density
\be
{\cal L} = \half(\dpsq{\phi}{\phi} - m^{2} \phi^{2}) + \half(\dpsq
{\sigma}{\sigma} - m_{\sigma}^{2}\sigma^2) - g\phi^2\sigma,
\label{lagrangian}
\ee
where $g$ is the $\phi$-$\sigma$ coupling constant.
\newsec{Formalism and PTIR}
\label{formal}
The Bethe-Salpeter equation in momentum space for a scalar-scalar bound 
state with scalar exchange is 
\be
\Phi(p,P)=-D(p_{1})D(p_{2})\int \frac{d^4q}{(2\pi)^4}\Phi(q,P)K(p,q;P),
\label{BSalg}
\ee
where $\Phi$ is the Bethe-Salpeter (BS) amplitude, and where $K$
is the scattering kernel, which contains information about the interactions
between the constituents of the bound state.

In Eq.~(\ref{BSalg}), $p_{i}$ is the four-momentum of the $i^{\rm th}$ 
constituent. We also define $p\equiv\eta_{2}p_{1}-\eta_{1}p_{2}$, which is 
the relative four-momentum of the two constituents, and $P\equiv p_{1}+p_{2}$ 
is the total 
four-momentum of the bound-state. The real positive numbers $\eta_{i}$ are 
arbitrary, with the only constraint being that $\eta_{1}+\eta_{2}=1$. 
In the case where the scalar constituents have equal mass, it is 
convenient to choose $\eta_{1}=\half=\eta_{2}$, and so henceforth these 
values of the $\eta_{i}$ will be used. This notation is used, for example, 
by Itzykson and Zuber~\cite{Zub}.

The quantities $D(p_i)$ are the propagators for the scalar constituents.
We will use free propagators here for simplicity, although we could include
arbitrary nonperturbative propagators by making use of their spectral 
representation~\cite{bd}:
\be
D(q^2)=-\frac{1}{m^2-q^2-i\epsilon}-\int_{(m+\mu)^2}^{\infty}d\alpha\:
     \frac{\rho_{\phi}(\alpha)}{\alpha-q^2-i\epsilon} ,
\label{spectral}
\ee
where $(m+\mu)$ is the invariant mass of the first threshold in $D$.
It is relatively straightforward to generalise the discussion below to
include $\rho_\phi(\alpha)\neq 0$.
We now redefine the scattering kernel $K$ such that $K(p,q;P)=i I(p,q;P)$,
and rewrite the momenta of the constituents in terms of the relative
momentum $p$ and the bound-state momentum $P$. The Bethe-Salpeter
equation becomes
\be
\Phi(p,P)=D(\halfP + p)D(\halfP - p)\int \frac{d^4q}{(2\pi)^4 i}\:\Phi(q,P)
          I(p,q;P).
\label{BSalgfin}
\ee
$I$ is the scattering kernel as referred to by Nakanishi in his 
review article~\cite{bsrev}.

In order to convert the BSE into a real integral equation, we will need to
use the Perturbation Theoretic Integral Representation (PTIR) for the
proper bound-state vertex and the scattering kernel~\cite{kw}. The bound-state
vertex may be represented as
\be
\Gamma^{[l,m]}(q,P)=\solidq\inthalf d\alpha\int_{-1}^{1}dz
\frac{\rho_n^{[l]}(\alpha,z)}{[\alpha-(q^2+zq\cdot P+\qtrP)-i\epsilon]^n}.
\label{basicvertex}
\ee
The weight function $\rho_n^{[l]}(\alpha,z)$ of the vertex has support
only for a finite region of the space spanned by the parameters $\alpha$
and $z$. A lower bound on the support is given by $\rho_n^{[l]}(\alpha,z)
=0$ unless
\bea
\lefteqn{\alpha\:\geq} 
    &\: & \:\max\left[\frac{1-z}{2}(m+\mu)^2 + \frac{1+z}{2}\left(m+\mu
          -\sqrt{P^2}\right)^2\right. , \non \\ 
    & & \left.\frac{1-z}{2}\left(m+\mu-\sqrt{P^2}\right)^2 
          + \frac{1+z}{2}(m+\mu)^2\right].  
\label{rhobarsupp}
\eea
The PTIR for the vertex (3-point) function was originally derived by 
Nakanishi~\cite{PTIR}; here we have also assumed that the vertex for 
bound-states
with non-zero angular momentum is given by the $s$-wave vertex multiplied
by the appropriate solid harmonic~\cite{nakanishi}. The solid harmonic is 
an $l^{\rm th}$ order polynomial of its arguments, and can be written as
\be
{\cal Y}_l^m(\vec{p})=|\vec{p}|^l Y_l^m(\hat{p}),
\label{solid}
\ee
with $Y_l^m(\hat{p})$ being the ordinary spherical harmonic for
angular momentum quantum numbers $l$ and $m$ and where $\hat{p}\equiv
\vec{p}/|\vec{p}|$ is a unit vector.

We have introduced a dummy parameter $n$, which will be of use in our
numerical work since larger values of $n$ produce smoother weight 
functions. The fact that $n$ is arbitrary can be seen by integrating
by parts with respect to $\alpha$; in this way weight functions for
different values of $n$ may be connected~\cite{kw}. 

We may use the PTIR for the bound-state vertex to derive the PTIR for the
Bethe-Salpeter amplitude, since the two are related via
\be
  \Phi(p,P)=iD(\halfP+p)i\Gamma(p,P)iD(\halfP-p).
\label{phigamma}
\ee
We proceed by absorbing the two free scalar propagators into the expression
for the vertex, Eq.~(\ref{basicvertex}), using the Feynman 
parametrisation~\cite{feyn}. After some algebra we obtain
\be
\Phi^{[l,m]}(p,P)=-i\solidp\int_{-1}^{1}dz
\intall d\alpha \frac{\varphi_{n}^{[l]}(\alpha,z)}{\left[m^2+\alpha-
(p^2+zp\cdot P +\qtrP)-i\epsilon\right]^{n+2}},
\label{pwBSE}
\ee
where the weight function for the BS amplitude, $\varphi_{n}^{[l]}
(\alpha,z)$, vanishes when
\be
\alpha < \min \left[0, (m+\mu+\sqrt{P^2})^2-m^2+\qtrP\right].
\label{weightsupp}
\ee 
To include the most general form of the scattering kernel in our derivation,
we use the PTIR for the kernel:
\bea
& &I(p,q;P)=\sum_{\rm ch}\inthalf d\gamma\int_{\Omega}d\vec{\xi}\:\non \\
& &\frac{\rho_{\rm ch}(\gamma,\vec{\xi})}{\gamma-(a_{\rm ch}q^2+b_{\rm ch}p
\cdot q+c_{\rm ch}p^2+d_{\rm ch}P^2+e_{\rm ch}q\cdot P+f_{\rm ch}p\cdot P)
-i\epsilon}, 
\label{generalkernel}
\eea
where the kernel parameters $\{a_{\rm ch},\ldots,f_{\rm ch}\}$ are 
linear combinations of the integration variables $\{\xi_1,\ldots,\xi_6\}$. 
Here we have defined, similarly to before, $q\equiv (q_1-q_2)/2$. The 
expression for the kernel contains a sum over three different channels,
labelled by $st$, $tu$ and $us$.

The support
properties of the kernel weight functions in each channel, $\rho_{\rm ch}$,
have been derived by Nakanishi~\cite{PTIR}. They will not concern us here
in our pure and generalised ladder treatments, since in both these cases
the kernel weight functions are simply products of delta functions.
\newsec{Derivation of Equations for Scalar Models}
\label{derivation}
Armed with Eqs.~(\ref{basicvertex}),~(\ref{pwBSE}) and~(\ref{generalkernel}), 
we 
can now derive real integral equations for the weight functions of the 
bound-state vertex and BS amplitude, both of which will be solved
numerically by iteration.

If we consider Eq.~(\ref{BSalgfin}), we note that it is necessary to combine
four factors using Feynman parametrisation. Having done this, we use the 
so-called ``self-reproducing'' property of the solid harmonics~\cite
{nakanishi,kw}:
\be
\int d\vec{q}\:F(|\vec{q}|^2)\solid{\vec{q}+\vec{p}}=
\solid{\vec{p}}\int d\vec{q}\:F(|\vec{q}|^2).
\label{self}
\ee
This allows us to perform the integral over the loop momentum $q$. Note
that in order for the loop-momentum integral to converge, we have the 
following restriction on the dummy parameter $n$:
\be
l<2n+2. \non
\ee
This arises from a simple power-counting argument, and tells us 
the values of $n$ that are valid for a particular partial wave. For
example, for $s$-wave solutions we may choose arbitrary positive $n$.

After the loop momentum integration and some algebraic manipulation,
we obtain the result
\be
\varphi_{n}^{[l]}(\ab,\zb)=\int_{0}^{\infty}d\alpha\int_{-1}^{1}dz
\:\varphi_{n}^{[l]}(\alpha,z){\cal K}_{n}^{[l]}(\ab,\zb;\alpha,z).
\label{mainphi}
\ee
We will omit the explicit expression for the kernel function ${\cal K}$
for the sake of brevity.

We can derive an equivalent equation to Eq.~(\ref{mainphi}), which has some
advantages for numerical solutions of the BSE. Inserting the PTIR for the 
vertex, Eq.~(\ref{basicvertex}), into the vertex BSE
\be
\Gamma(p,P)=\int\frac{d^4q}{(2\pi)^4}D(\halfP+q)D(\halfP-q)\Gamma(q,P)I(p,q;P),
\label{vertexBSE}
\ee
and once again using the self-reproducing property of the solid 
harmonics to perform the loop integral, we obtain the result
\be
\rho_{n}^{[l]}(\ab,\zb)=\inthalf d\alpha\int_{-1}^{1}dz
\:\rho_{n}^{[l]}(\alpha,z)\bar{{\cal K}}_{n}^{[l]}(\ab,\zb;\alpha,z).
\label{mainrho}
\ee
Once again the explicit form of ${\bar{\cal K}}$~\cite{kkt} will be 
omitted for brevity.

Once we solve for either $\varphi_{n}^{[l]}(\alpha,z)$ for the BS amplitude or 
$\rho_{n}^{[l]}(\alpha,z)$ for the vertex, 
we can evaluate both the BS amplitude 
and the vertex for any momenta $p$ and $P$ without solving any additional 
integral equation.
It should be also mentioned that our approach is completely independent of the 
choice of inertial frame, so that no Lorentz boost is necessary to obtain 
the BS vertex for a moving bound state. These are very useful
properties when applying the BS vertex to calculations of physical
processes involving bound states.

\newsec{Numerical Solutions and Results}
\label{numres}
For our numerical study, we have specialised to the case of $s$-wave
($l=0$) bound states interacting via a pure ladder kernel,
\be
I(p,q;P)=\frac{g^2}{m_\sigma^2-(p-q)^2-\ie},
\ee
which corresponds to the following fixed set of kernel parameters:
$a_{st}=1=c_{st}$, $b_{st}=-2$, $\gamma=m_\sigma^2\:$ for the $st$-channel,
with the weight functions for the $tu$ and $us$ channels vanishing. We have
also solved the BSE for a sum of the pure ladder kernel and a 
generalised ladder term, which has non-vanishing fixed values for the
kernel parameters $d_{\rm ch}$, $e_{\rm ch}$ and $f_{\rm ch}$~\cite{kw}. 

For numerical solution of Eqs.~(\ref{mainphi}) and (\ref{mainrho}), it is
convenient to define an ``eigenvalue'' parameter $\lambda$, with
$\lambda$ being defined in terms of the coupling as $\lambda\equiv g^2/(4\pi)
^2$,
and writing the kernel functions ${\cal K}$ and $\bar{\cal K}$ as 
$\lambda K$ and $\lambda\bar{K}$, respectively.
This having been done, we solve the BSE with the new kernel functions
$K$ and $\bar{K}$ as an eigenvalue equation,
by iteration. Our approach enables us to obtain the eigenvalue as
a function of the bound-state mass squared, as well as the weight
functions for the BS amplitude and the vertex. 

The BSE for the amplitude has been solved in a previous work~\cite{kw},
with accuracy of the order of a few parts in 100. We have also solved
here the BSE for the vertex, which has a simpler structure. These 
vertex solutions,
obtained for $n=2$ using an optimised grid and sophisticated integrator 
and interpolator, have an
accuracy of approximately 1 part in $10^4$.
Our results for the eigenvalue for the
case where $m_\sigma/m=0.5$, $m_\sigma$ being the mass of the
exchanged $\sigma$-particle, are shown in Table~1, and are
plotted in Fig.~1.
Some examples of vertex weight functions for various values of
the bound state mass are shown in Fig.~2. What is actually plotted
for convenience is the rescaled weight function,
$\rho\equiv\rho^{[\ell]}_n/\alpha^n$,
where here $n=2$ and $\ell=0$.  The parameter
$\eta$ represents the ``fraction of binding'', $\eta\equiv M/2m$, 
$M\equiv\sqrt{P^2}$ 
being the bound state mass.
\begin{table}[hbt]
\begin{center}
\begin{tabular}{|l|r|r||l|r|r|} \hline
$\eta$ & $\lambda_{E}$ & $\lambda_{M}$ & $\eta$ & $\lambda_{E}$ & 
$\lambda_{M}$ \\ \hline 
0 & 2.5658    & 2.56615
& 0.9 & 1.0349    & 1.03497 \\ \hline
0.2 & 2.4984    & 2.49883 & 0.95 & -  & 0.79528 \\ \hline
0.4 & 2.2933    & 2.29371 & 0.99 & 0.5167    & 0.51684 \\ \hline
0.6 & 1.9398    & 1.94020 & 0.999 & 0.3852    & 0.38530 \\ \hline
0.8 & 1.4055    & 1.40560 & 1 & 0.3296    & - \\ \hline
\end{tabular}

\parbox{130mm}{\caption
{Eigenvalues for the 
Wick-rotated ($\lambda_E$) and Minkowski-space ($\lambda_{M}$) solutions 
of the Bethe-Salpeter Equation. The Wick-rotated values are from
Linden and Mitter\protect\cite{lm}.}}
\end{center}
\end{table}

\newsec{Conclusions and Outlook}
\label{conc}
We have obtained numerical solutions of the Minkowski space BSE for 
scalar-scalar bound states in the pure ladder model. Our results agree
exactly with those obtained in the Euclidean space treatment of Linden
and Mitter~\cite{lm}. Our technique can be generalised to arbitrary
scalar-scalar bound states, given that we know expressions for the
kernel weight functions $\rho_{\rm ch}$ and the propagator spectral
function $\rho(s)$. The key to our approach is to convert the BSE from
an integral equation involving complex distributions into one involving 
functions,
which is numerically soluble.

Our numerical solutions have been obtained for some simple choices
of kernel weight function. It remains for us to carry out
systematic studies for non-ladder kernels, orbital excitations, and 
comparison with other (approximate) methods such as solutions for
separable kernels~\cite{kkt}.
It will also be desirable to include spectral functions for the
constituent particle propagators in a more sophisticated treatment
{\em e.g.}, the ``dressed ladder'' kernel, which involves the simultaneous
solution of self-energy Dyson-Schwinger equations and the Bethe-Salpeter
equation.

To date the spinor-spinor and scalar-spinor BSEs have been solved only in
the ladder approximation~\cite{seto}. It will be important to attempt to 
formulate an approach
for fermions similar to the one outlined here, so that we may solve the
BSE involving spinors for more ``realistic'' scattering kernels. We 
anticipate that the most challenging aspect of this will be generalising 
the PTIR to particles with non-zero spin. 
Since we would like to study the bound-state problem
in QCD, for example the fermion-antifermion BSE for mesons, we also
need to resolve the problem of incorporating confinement into the PTIR,
and of including derivative couplings.

\newpage
\begin{center}
CAPTIONS FOR FIGURES
\end{center}
Figure 1: The bound state spectrum for equal constituent masses and
an exchange mass of $m_\sigma=0.5 m$.

Figure 2: The rescaled weight function
$\rho\equiv\rho_2^{[0]}/\alpha^2$ of the bound-state
vertex for various values of $\eta$ in the pure ladder limit. $\eta=0$
corresponds to the case of a massless (i.e., Goldstone-like) bound state.

\newpage
\begin{figure}[htb]
  \centering{\
     \epsfig{angle=0,figure=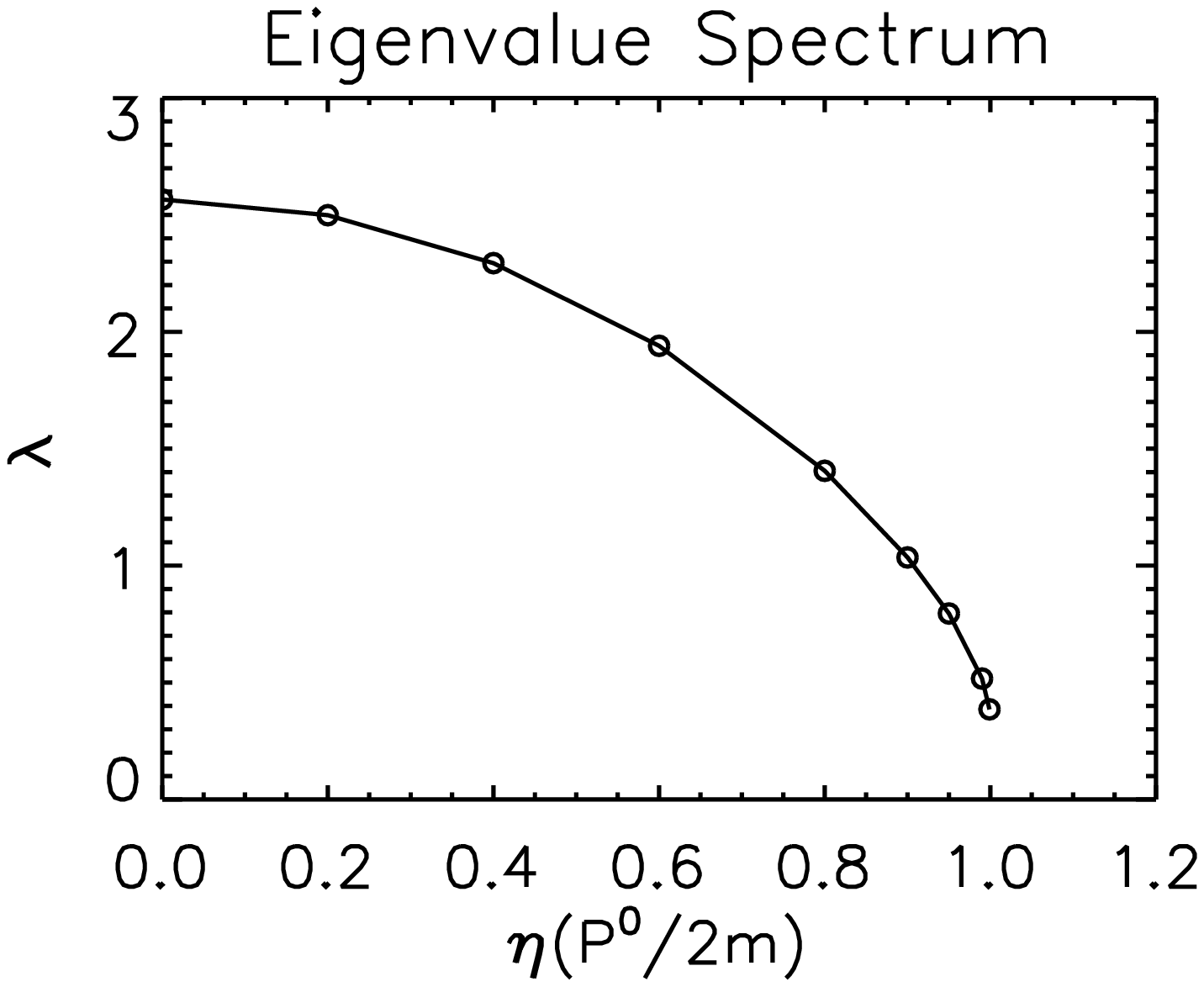,height=10.5cm}  }
\label{spectrumplot}
\end{figure}

\newpage
\begin{figure}[htb]
  \centering{\
     \epsfig{angle=0,figure=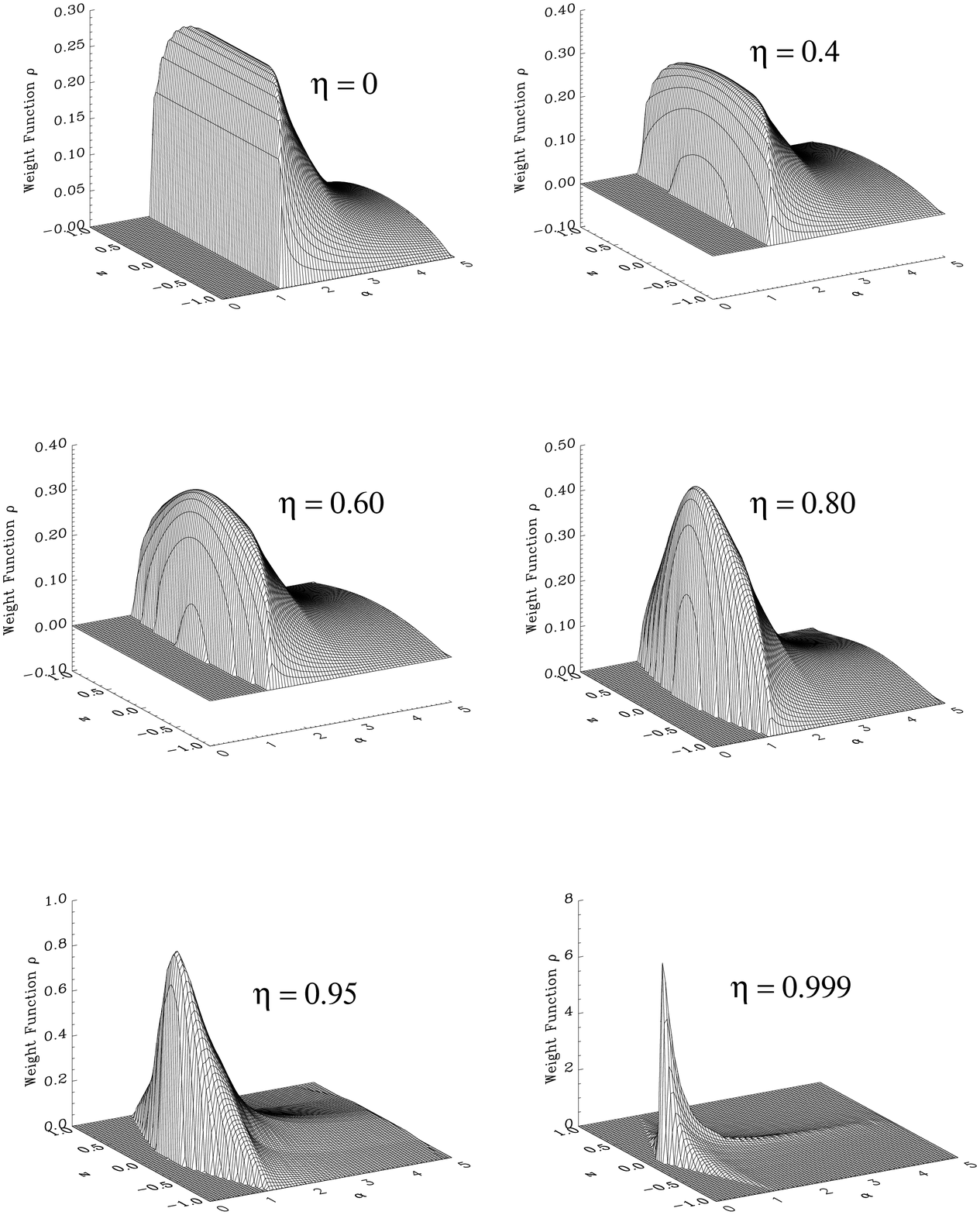,height=19.5cm}  }
\label{weightfunc}
\end{figure}
\end{document}